%% file: draft.tex
\documentclass[preprint, 12pt]{elsarticle}

\usepackage{latexsym}
\usepackage{graphicx}
\usepackage{color}
\usepackage{amsmath}  
\usepackage{amssymb}
\usepackage{longtable}

\usepackage{hyperref}
\usepackage{natbib}

\begin{document}

\begin{frontmatter}

\title{The STELLA Apparatus for Particle-Gamma Coincidence Fusion Measurements with Nanosecond Timing}


\include{stella_collab}


\begin{abstract}
  The STELLA (STELlar LAboratory) experimental station for the measurement of deep sub-barrier light heavy-ion fusion cross sections has been installed at the Androm\`{e}de accelerator at the Institut de Physique Nucl\'{e}aire, Orsay (\textit{France}). The setup is designed for the direct experimental determination of heavy-ion fusion cross sections as low as tens of picobarn. The detection concept is based on the coincident measurement of emitted gamma rays with the UK~FATIMA (FAst TIMing Array) and evaporated charged particles using a silicon detector array. Key developments relevant to reaching the extreme sub-barrier fusion region are a rotating target mechanism to sustain beam intensities above $10\mu$A, an ultra-high vacuum of 10$^{-8}$~mbar to prevent carbon built-up and gamma charged-particle timing in the order of nanoseconds sufficient to separate proton and alpha particles.
\end{abstract}

\end{frontmatter}


\section{Introduction}
\label{sec:intro}

Heavy-ion fusion reactions involving $^{12}$C and $^{16}$O nuclei such as the $^{12}$C+$^{12}$C reaction play a key role in the evolution of massive stars and in explosive astrophysical scenarios such as type Ia supernovae and super-bursts in binary systems. Since the 1950s the $^{12}$C+$^{12}$C system was well known to exhibit strongly resonant behaviour~\cite{bromley1960} which also manifests in the fusion cross-section with prominent resonances, at energies ranging from a few MeV per nucleon, down to the Coulomb barrier and below~\cite{back2014}. Such resonances have been attributed to the formation of long-lived $^{12}$C+$^{12}$C molecular configurations.

The presence of these resonances will inevitably have a strong impact on the carbon burning reaction rates under the different astrophysical scenarios. Direct cross-section measurements are therefore needed down into the Gamow window corresponding to carbon burning in massive stars. These experiments are hugely challenging as the relevant cross-sections are well below the nanobarn level. Reaction rates presently rely on extrapolations of cross-section data from higher energies. These data are largely based either on the detection of evaporated charged particles ($\alpha$, p) or the characteristic gamma decay of the $\alpha$ and p evaporation residues $^{23}$Na and $^{20}$Ne~\cite{aguilera2007}. The former technique suffers from the presence of low-level deuterium contamination in the carbon target as the reaction $^{12}$C(d,p) has a large cross section and the resulting protons are at similar energies to the far weaker evaporated charged particles from $^{12}$C+$^{12}$C fusion. Gamma-ray detection is challenging at the level of the cross-sections of interest from the point of view of discriminating signal from background. A clear way to achieve a system with strong background suppression is to measure coincidences between evaporated charged particles and their associated gamma rays since this is a unique signature. This technique has been realised by Jiang {\em et al.}~\cite{jiang2012, jiang2018} using the Gammasphere germanium detector array and an annular silicon detector array at Argonne National Laboratory. Their initial results are very promising but the full potential of the technique is limited by the available beam currents (of the order of 100 pnA) and the potential running period (of the order of one week). To extend this approach to the energies of astrophysics interest, such experiments will need beam currents in the microampere range and extended running periods of many weeks duration. This is the challenge addressed by the STELar LAboratory described in this paper. The key elements of STELLA are:

\begin{itemize}
  \item rotating targets which can sustain high beam intensities,
  \item high-efficiency particle and gamma-ray detection systems, and
  \item employment of a coincident technique which allows the extraction of the relevant fusion signal from the dominant background.
\end{itemize}

\section{Apparatus}
\label{sec:exp}

\begin{figure}[htp]
  \center
  \includegraphics[width=.8\textwidth]{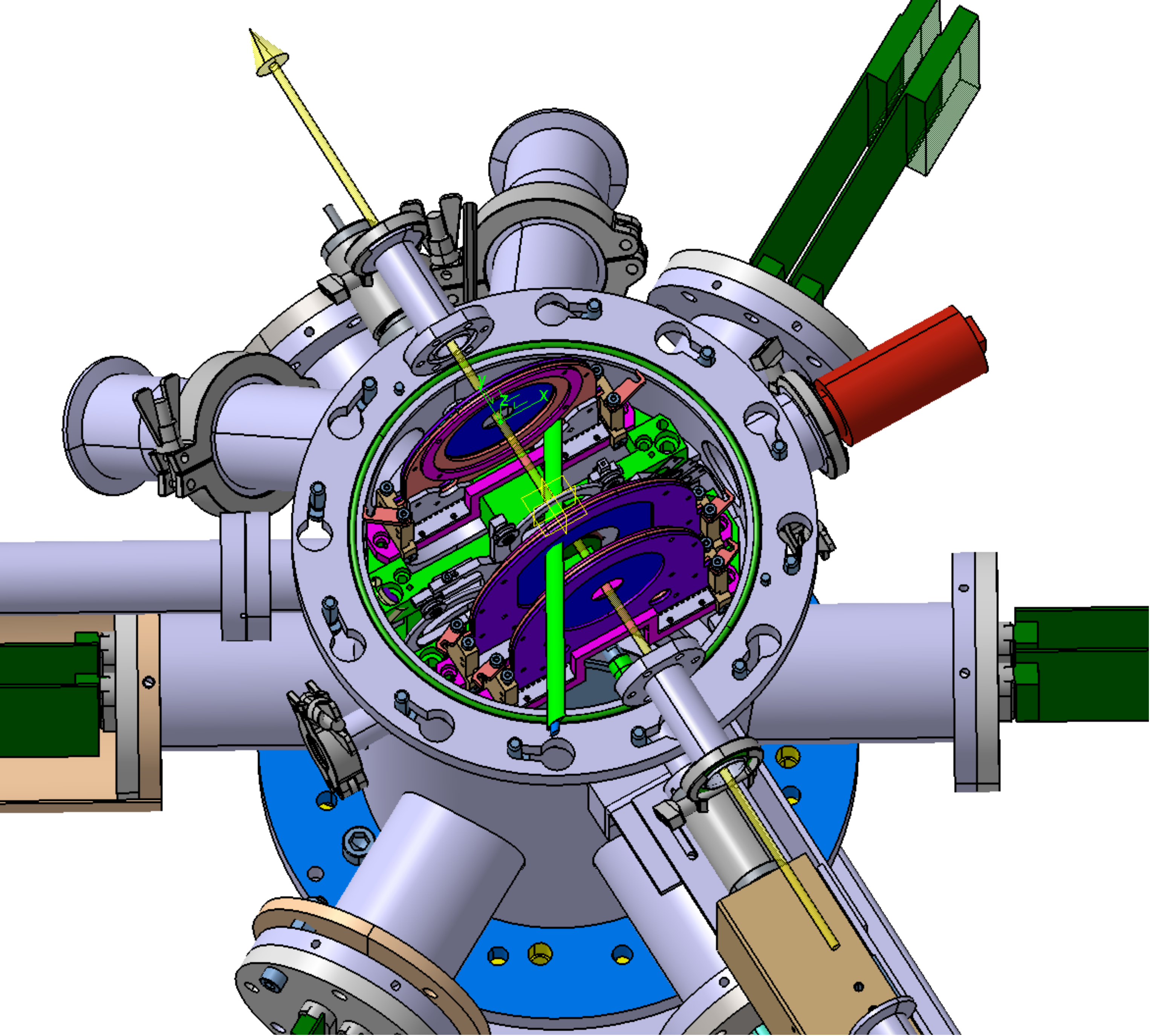}
  \caption{[color online] View into the target chamber that is mounted on top of the cryogenic pump and that is closed by a this Al dome. The annular particle detectors shown in dark blue are aligned along the beam axis around the target at the center of the reaction chamber. The extensions serve as feed-throughs for detector signals and host two surface barrier silicon monitor detectors at 45$^{\circ}$ with respect to the beam line.}
  \label{fig:top_cha}
\end{figure}

The scattering chamber of the STELLA system is presented in Figure~\ref{fig:top_cha}.
The chamber contains several annular DSSSDs (Double-Sided Silicon-Strip-Detector), described in detail in section~\ref{sec:dsssd}, for high-efficiency particle measurements. The detectors are aligned along the beam axis around the target (see section~\ref{sec:rt}) at the center of the reaction chamber. All support structures and signal cables are directed towards the bottom of the chamber where a cryogenic ultra-high vacuum pump is located providing a vacuum of $10^{-8}$~mbar. The gamma-ray detectors comprising an array of lanthanum bromide (LaBr$_{3}$(Ce)) scintillators are supported from above and surround the 2.5~mm thick aluminum dome-shaped target chamber with a diameter of 20~cm. The gamma-detection array is introduced in section~\ref{sec:labr3}. All data are time-stamped with sampling times of 1~ns and 8~ns, respectively, for gamma ray and charged-particle detection. The synchronization of coincident gamma-particle events in $^{12}$C+$^{12}$C fusion reactions as well as a first background reduction estimate is discussed in section~\ref{sec:daq}. Two surface barrier silicon detectors for the measurement of scattered beam particles as well as a Faraday integrator are used for the precise determination of the beam intensity during the measurements. The beam particle monitors are located 23~cm from the target in extensions of the reaction chamber that form an angle of 45$^{\circ}$ with the beam line.

\subsection{Rotating Target}
\label{sec:rt}

\begin{figure}[htp]
  \center
  \includegraphics[width=.6\textwidth]{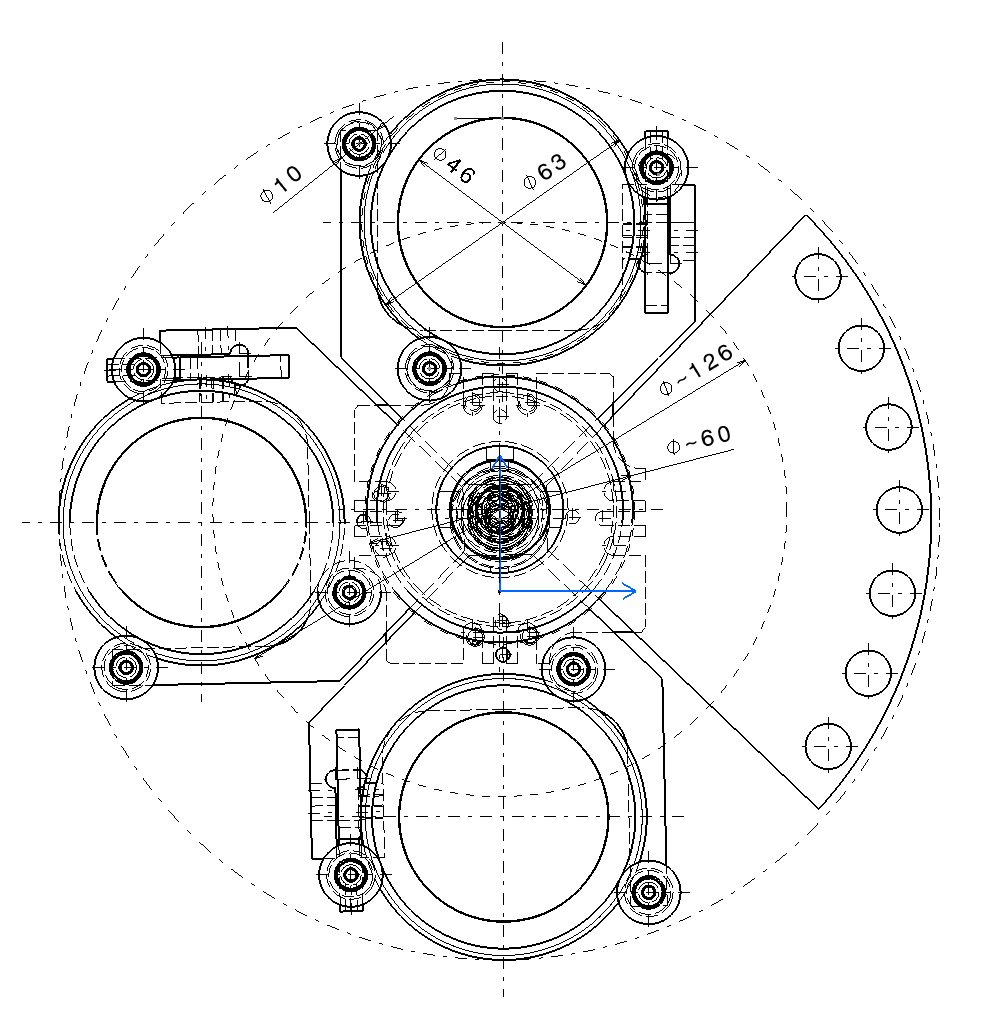}
  \caption{Front view of the target wheel where the uppermost quadrant is exposed to the beam. The wheel hosts a quadrant with seven fixed target slots (on the right) and three rotating targets with a diameter of 6.3~cm. The central wheel transmits the rotation from the external motor to the target frames.}
  \label{fig:xs_tar}
\end{figure}

With $^{12}$C beam intensities in the order of p$\mu{}$A and a beam spot diameter of 2~mm, a heat input of Watts may be estimated based on energy loss for targets of a few tens $\mu{}$g/cm$^{2}$ thickness. In order to avoid breaking of the targets, it was necessary to develop the rotating target mechanism displayed in Figure~\ref{fig:xs_tar}.
It is designed to increase the effective beam spot size to distribute the thermal load. The wheel hosts three rotating target frames and seven slots for fixed target experiments. A magnetic feed-through connects the rotation-driving motor outside the vacuum with the central wheel to spin \textit{via} friction with the target frame bearing in contact. This bearing transmits the rotation to the rotating target frames, each with a diameter of 6.3~cm. In total, three bearings guide each target frame. The axis of the target revolver mechanism used to change the target is slightly off the target rotation axis. In this way, only the uppermost target can spin, because the other bearings are not in contact with the drive shaft.

The layout is optimized for heat dissipation using the MATHCAD15$^{\textregistered}$ package where the temperature distribution at the beam spot position is calculated solving the heat-flow equation with a radiative heat loss term. This follows the Stefan-Boltzmann law at high excess temperatures and the net-radiative heat loss over time is obtained with $P_{rad} = \epsilon{}\sigma{}S\cdot{}(T^{4} - T^{4}_{s})$, where the emissivity $\epsilon = 0.8$, $\sigma{}$ is the Stefan-Boltzmann constant, $S$ is the surface area during one turn of the target, $T$ the temperature of the environment is $20 ^{\circ}$C and $T_{s}$ is the target foil temperature.

\begin{figure}[htp]
  \center
  \includegraphics[width=.7\textwidth]{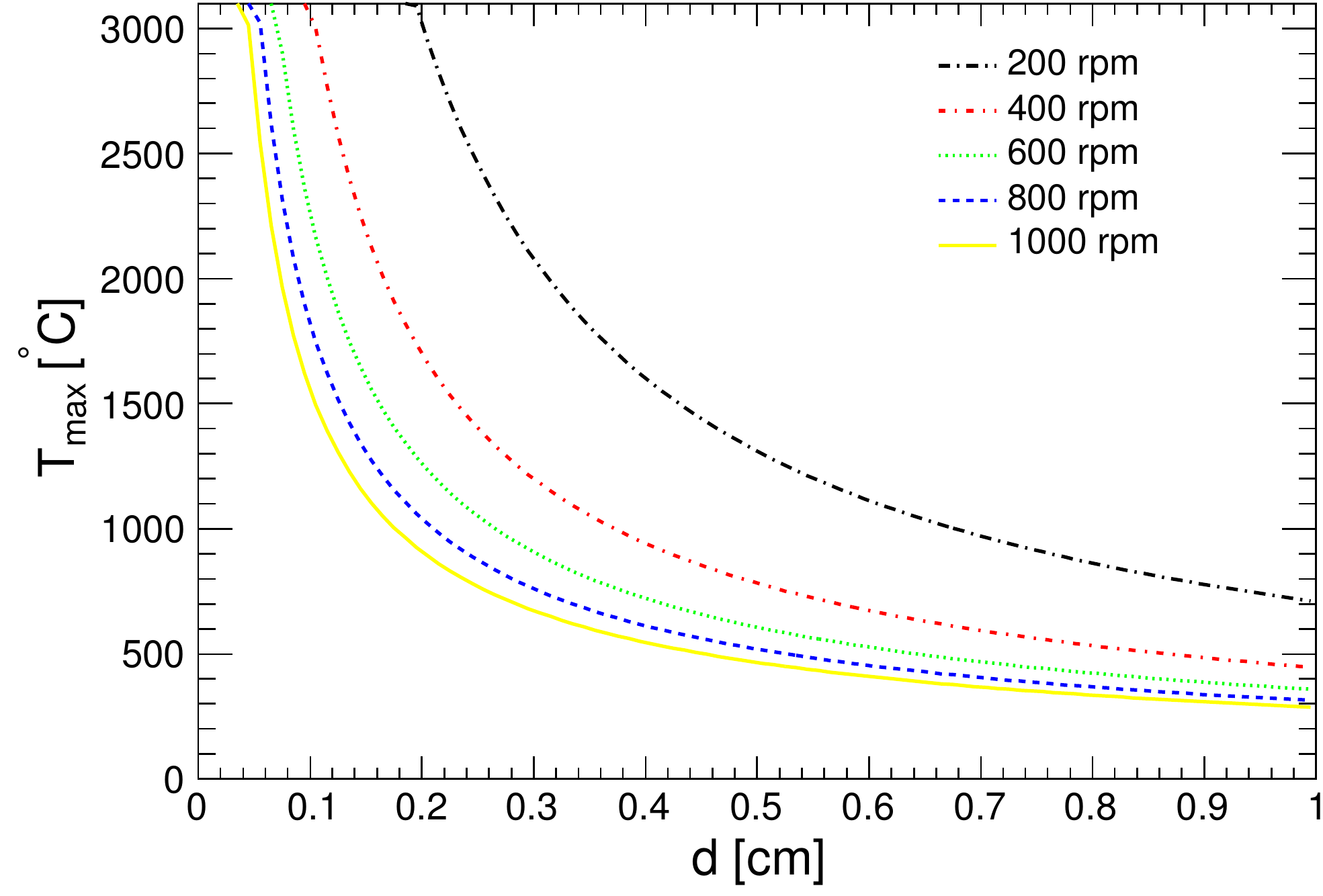}
  \caption{[color online] Dependency of the maximum temperature $T_{max}$ from beam spot diameter $d$ at a heat input power of 2~W for different rotation velocities.}
  \label{fig:dT_dd}
\end{figure}

The voxels of the target material are heated when exposed to the beam and they cool \textit{via} radiation when off the beam axis during the rotation. Taking into account these effects, the time-dependent profile of the target temperature may be calculated per turn of the target frame. The resulting temperature converges towards a maximum $T_{max}$ within seconds for the chosen parameters with a saw-tooth like cooling modulation $\Delta{}T$ between two heating pulses. The dynamics are mainly dependent on the beam spot size, the radius of the beam track on the target, and the rotation velocity at a given beam intensity. An example of the multi-parameter study is given in Table~\ref{tab:temp_vs_spot_1000}
\begin{table}[hpt]
  \centering
  \begin{tabular}{cc c cc c cc}
    \hline\hline
    $d$ [mm] & $T_{max}$ [C$^{\circ}$] & & $d$ [mm] & $T_{max}$ [C$^{\circ}$] & & $d$ [mm] & $T_{max}$ [C$^{\circ}$] \\
    \hline
    \multicolumn{2}{c}{$P=1$~W} & & \multicolumn{2}{c}{$P=2$~W} & & \multicolumn{2}{c}{$P=3$~W} \\
    \cline{1-2} \cline {4-5} \cline{7-8}
    \multicolumn{2}{c}{} & & \multicolumn{2}{c}{} & & \multicolumn{2}{c}{} \\
    2 & 550 & & 2 & 920 & & 2 & 1240 \\
    3 & 410 & & 3 & 670 & & 3 & 910 \\
    4 & 340 & & 4 & 550 & & 4 & 730 \\
    5 & 290 & & 5 & 470 & & 5 & 620 \\
    \hline\hline
  \end{tabular}
  \caption[why this?]{Temperature dependence $T_{max}$ from the beam spot diameter $d$ for various heat input power $P$ at a rotation velocity of 1000~rpm.}
  \label{tab:temp_vs_spot_1000}
\end{table}
for a rotation speed of 1000~rpm. The maximum temperature $T_{max}$ is listed depending on the beam spot diameter $d$ for various heat input power $P$. A higher rotation speed leads to more efficient cooling as the heat input per voxel decreases. The radiative cooling is most efficient right after the heat is deposited and the shortened rotation cycle has a negligible impact. The dependency of maximum temperature on the beam spot diameter is presented in Figure~\ref{fig:dT_dd}
for different rotation velocities at a heat input power of 2~W. It can be seen that for small beam diameters a rotation with 1000~rpm is necessary to keep the target material below a temperature of 1000$^{\circ}$C. This value serves as an empirical benchmark to ensure that the material properties of the carbon target remain unchanged during irradiation. With a target frame diameter of 6.3~cm, the trajectory of a beam spot close to the frame becomes around 14~cm long which is sufficient to keep the maximum temperature of the target material below the benchmark value.

\subsection{Charged Particle Detection}
\label{sec:dsssd}

Light charged particles from the reaction are detected by a set of annular S1- and S3-type DSSSD (Double-Sided Silicon Strip Detector) based on chips manufactured by Micron Semiconductor Ltd., where the S1~(S3) chips are segmented in 16~(24) rings on the junction side and 16~(32) sectors on the ohmic side. In the design (see Figure~\ref{fig:top_cha}) developed by the IPHC Mechanics and Microtechnique Department at CNRS Strasbourg, the chips are sandwiched between low outgassing RO4003C Rogers$^{\textregistered}$ ceramics which serve to replace a regular PCB in its role of providing detector polarization and signal readout. The same design permits to fit in chips of 500~$\mu$m (S1/S3) or 1000~$\mu$m (S3) thickness where the incomplete rings of S1 are connected to a closed circle with an adapted PCB cabling.

The signal connection to the front-end electronics is via a series of contacts at the base of the PCB connected via spring-like pins on the detector support inside the reaction chamber. This connection system is integrated into the vertical slots of the sliding system for the PCBs that are kept in position with a precision better than 1~mm using clamps. Low-outgassing Kapton$^{\textregistered}$ insulated cables feed the electronics signals into sets of MPR-16D differential Mesytec$^{\textregistered}$ preamplifier cards outside the reaction chamber before processing towards the digitizers.

Integrated aluminum absorber foils in front of the silicon detectors protect them from delta electrons and radiation damage from scattered beam particles under experimental conditions. The thickness of the foils is adapted to minimize the degradation of the proton and alpha-particle energies. The system is grounded to mitigate the effects of charging. The junction side of the DSSSD is biased with a negative potential, while the ohmic side facing the target is grounded, thus guarding against damage due to possible sparking between the protecting aluminum foil and the detector surface.

\begin{figure}[htp]
  \center
  \includegraphics[width=.7\textwidth]{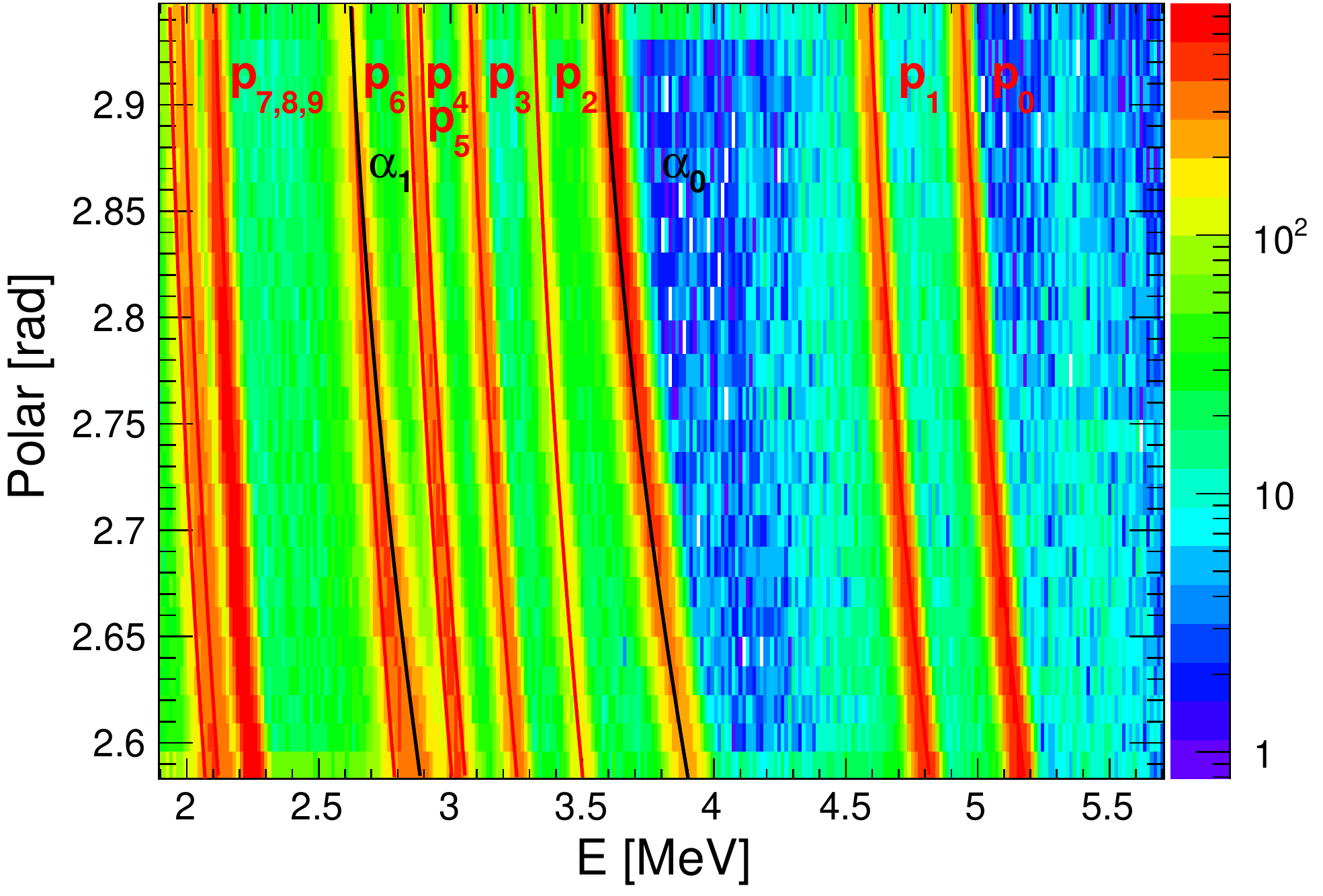}
  \caption{[color online] Angular distribution of protons and alphas associated with various excitation levels $i$ of the corresponding fusion evaporation nucleus $^{23}$Na ($p_{i}$) and $^{20}$Ne ($\alpha_{i}$), respectively, for the S3 in backward direction at a beam energy of 11~MeV.}
  \label{fig:dsssd_ene_vs_ang}
\end{figure}

The annular charged-particle detectors, placed along the beam axis, are displayed in blue in Figure~\ref{fig:top_cha} in the top view of the scattering chamber. Upstream a pair of S3 and S1 detectors located 5.6~cm and 3.1~cm, respectively, from the target. The relative positioning is chosen to avoid shadowing of the target vertex. At the same time, the angular coverage is maximized. Downstream, an S3 is at around 6~cm from the target. In this configuration, the angular acceptance is 30$\%$ of the solid angle. The angular coverage per strip ranges from 10.0~mrad (outer ring S3) to 27.8~mrad (inner ring S1) due to the compact geometry of the system. The relative energy resolution obtained with the $\alpha$-emitter $^{239}$Pu is 0.5$\%$ FWHM at 5154~keV. An energy spectrum with light charged particles from the $^{12}$C+$^{12}$C reaction at a beam energy of 11~MeV is shown in Figure~\ref{fig:dsssd_ene_vs_ang},
for the S3 detector in backward direction. In the angular range, various bands with protons and alphas, where the index stands for the excitation level of the associated daughter nucleus $^{23}$Na and $^{20}$Ne, respectively, are labeled. The solid lines are kinematics calculations for the emitted particles. The angular resolution is sufficient to cleanly distinguish the exit channels. For clean separation of protons and alphas, additional selection criteria based on timing are used (see section~\ref{sec:gp_coinc}).

\subsection{$LaBr_{3}$ Array}
\label{sec:labr3}

\begin{figure}
  \centering
  \includegraphics[width=0.5\textwidth]{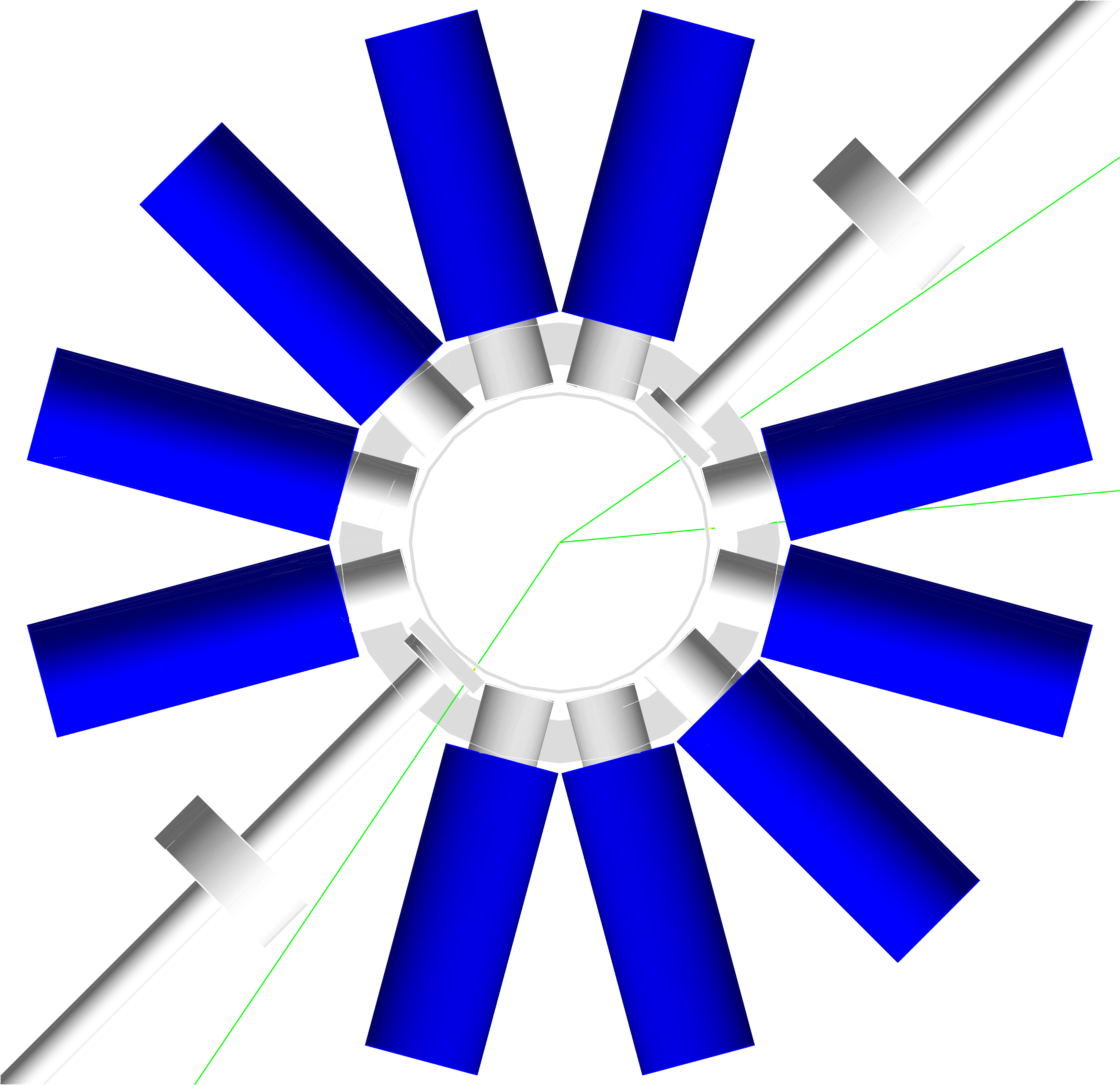}
  \caption{[color online] Top view onto the lowest ring with ten LaBr$_{3}$ detectors in a spherical configuration where the crystals face the target position. The scattering chamber and the dome aren't shown for simplicity.}
  \label{fig:sim_sphere}
\end{figure}

\begin{figure}
  \centering
  \includegraphics[width=0.5\textwidth]{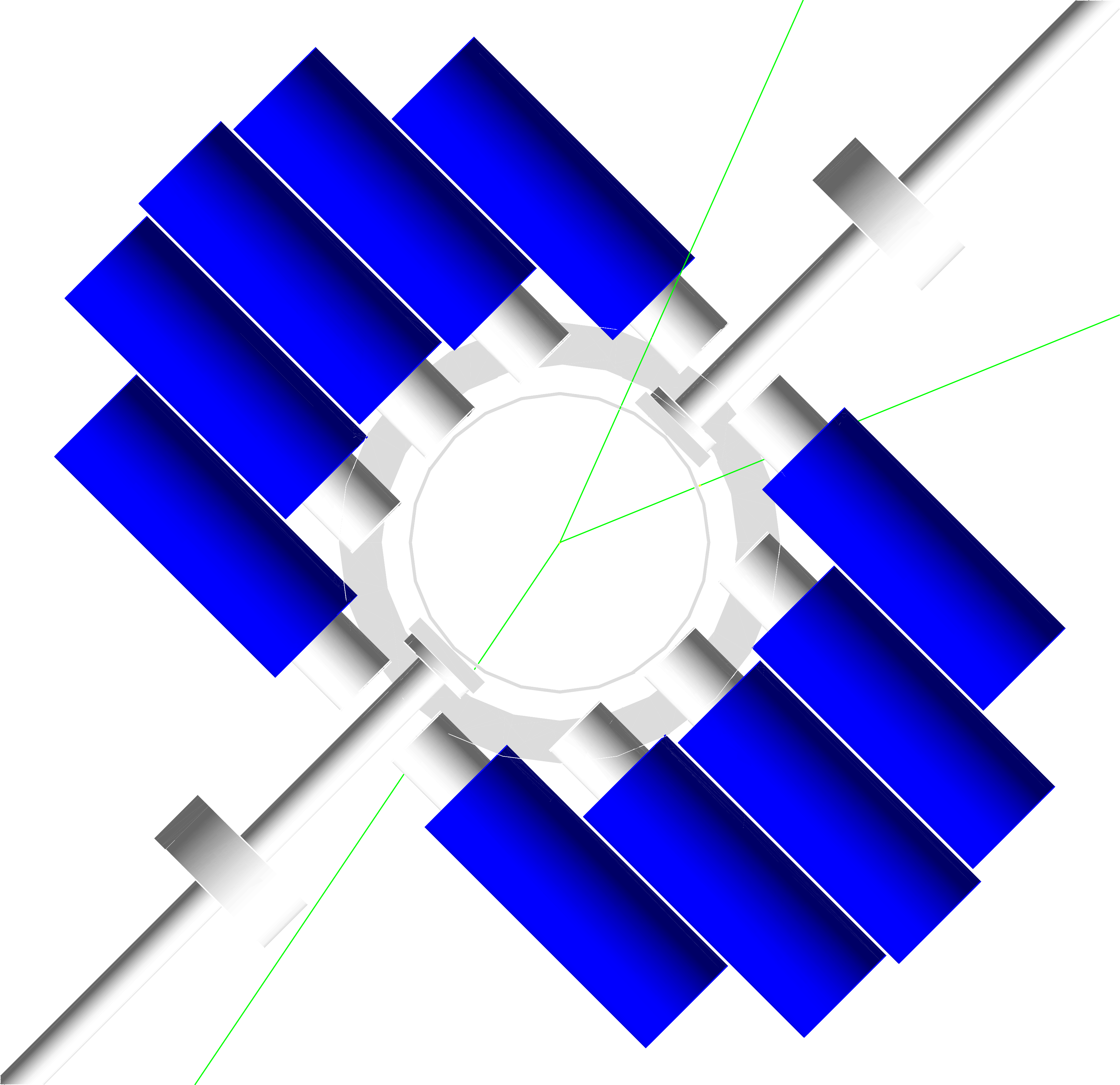}
  \caption{[color online] Top view onto the lowest line with ten LaBr$_{3}$ detectors in a cylindrical configuration where the crystals face the beam line. The scattering chamber and the dome aren’t shown for simplicity.}
    \label{fig:sim_cylinder}
\end{figure}

\begin{figure}[htp]
  \center
  \includegraphics[width=.7\textwidth]{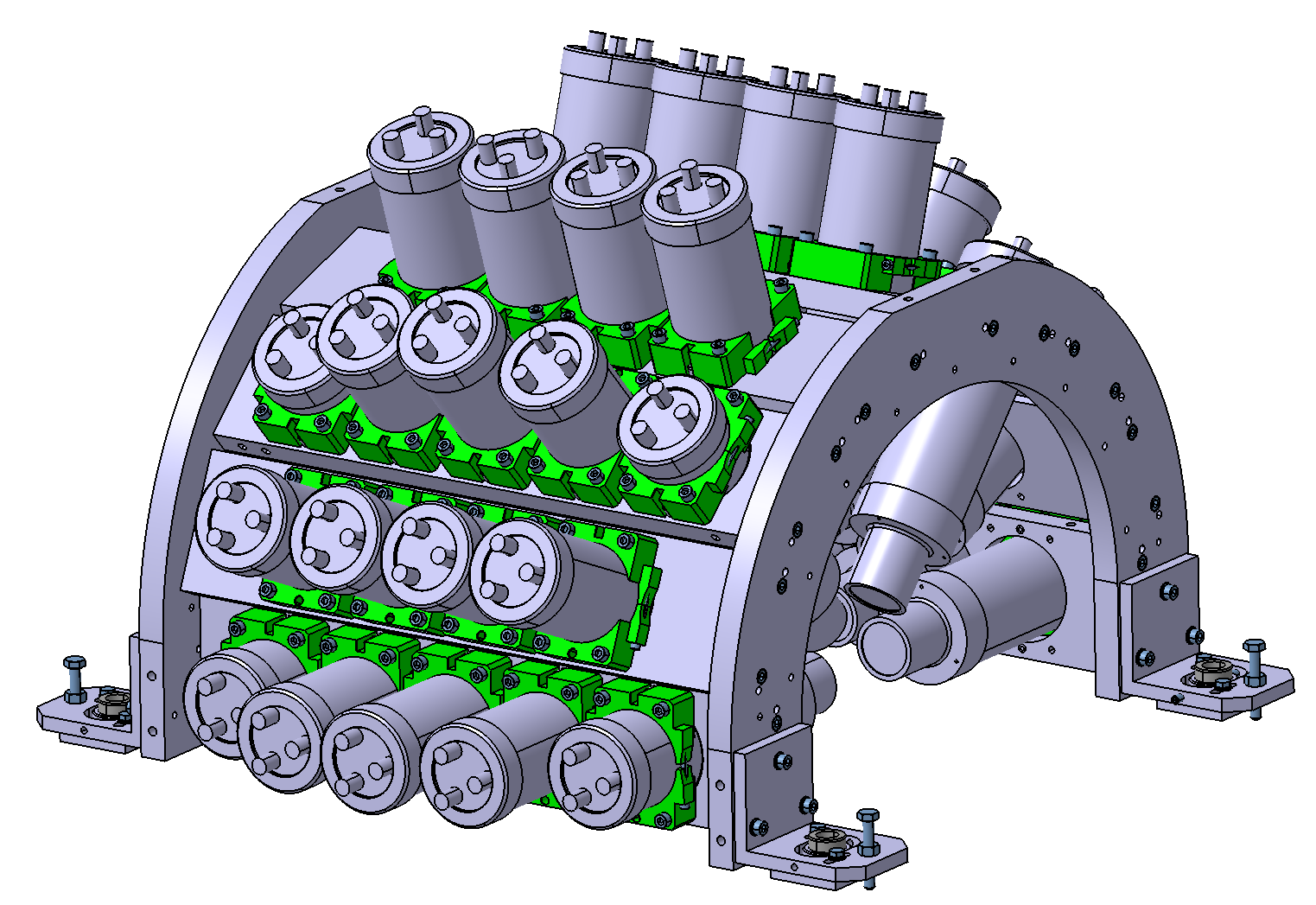}
  \caption{[color online] Final design of an array 36~LaBr$_{3}$ crystals housed in the 2'' tubes, read out by 3'' photo multipliers. Groups of five or six detectors are organized in shelves that are oriented towards the beam line.}
  \label{fig:gammaArray}
\end{figure}

Gamma rays from the de-excitation of fusion-evaporation residues are detected with an array of 36 LaBr$_{3}$ crystals of 1.5'' diameter and 2'' length from the UK~FATIMA collaboration~\cite{roberts2014, regan2015, shearman2017}. These detectors have sub-nanosecond timing resolution and an energy resolution of 3$\%$~FWHM at 1333~keV determined with a $^{60}$Co source. The detectors are positioned in a frame, supported from above. They can be lowered on rails to achieve a close packing of the top of the 2.5~mm thick aluminum dome of the scattering chamber. Their final approach to the chamber surface is guided by a set of pins for a placement precision better than 1~mm. Several detector configurations were analyzed in a simulation study focusing on full-energy peak detection efficiency. The lowest ring of a spherical as well as the lowest line of a cylindrical assembly are presented in Figure~\ref{fig:sim_sphere} 
and Figure~\ref{fig:sim_cylinder} 
where the positioning is optimized to the closest approach of the chamber at the center. Taken together with the feasibility, this study strongly favoured the final design in the cylindrical alignment shown in Figure~\ref{fig:gammaArray} and described in detail in Ref.~\cite{heine2016}.
In this configuration, the geometrical acceptance is 23$\%$ of the solid angle, with the full-energy peak gamma-ray detection-efficiency listed in Table~\ref{tab:gam_detEff}.
\begin{table}[hpt]
  \centering
  \begin{tabular}{c | cc cc cc cc cc}
    \hline\hline
    E [MeV] & 0.01 & 0.44 & 1.0 & 1.63 & 2.0 & 3.0 & 4.0 & 5.0 & 6.0 & 7.0 \\
    \hline
    $\epsilon_{sing}$ [$\%$] & 23.1 & 8.0 & 3.5 & 2.2 & 1.8 & 1.1 & 0.7 & 0.5 & 0.4 & 0.4 \\
    $\epsilon_{sum}$ [$\%$]  & 23.1 & 8.6 & 4.1 & 2.6 & 2.1 & 1.4 & 1.0 & 0.7 & 0.6 & 0.5 \\
    \hline\hline
  \end{tabular}
  \caption{Full-energy peak gamma detection efficiency $\epsilon{}$ of 36~LaBr$_{3}$ detectors in percent. The efficiency obtained from the analysis of single detector spectra $\epsilon_{sing}$ is compared to the value reflecting the total energy deposit in the array $\epsilon_{sum}$.}
  \label{tab:gam_detEff}
\end{table}
The fraction of energy entries in multiple detectors reflects in $\epsilon_{sum}$ and the analysis of the total energy deposit is more important towards higher gamma energies.

\begin{figure}[htp]
  \center
  \includegraphics[width=.7\textwidth]{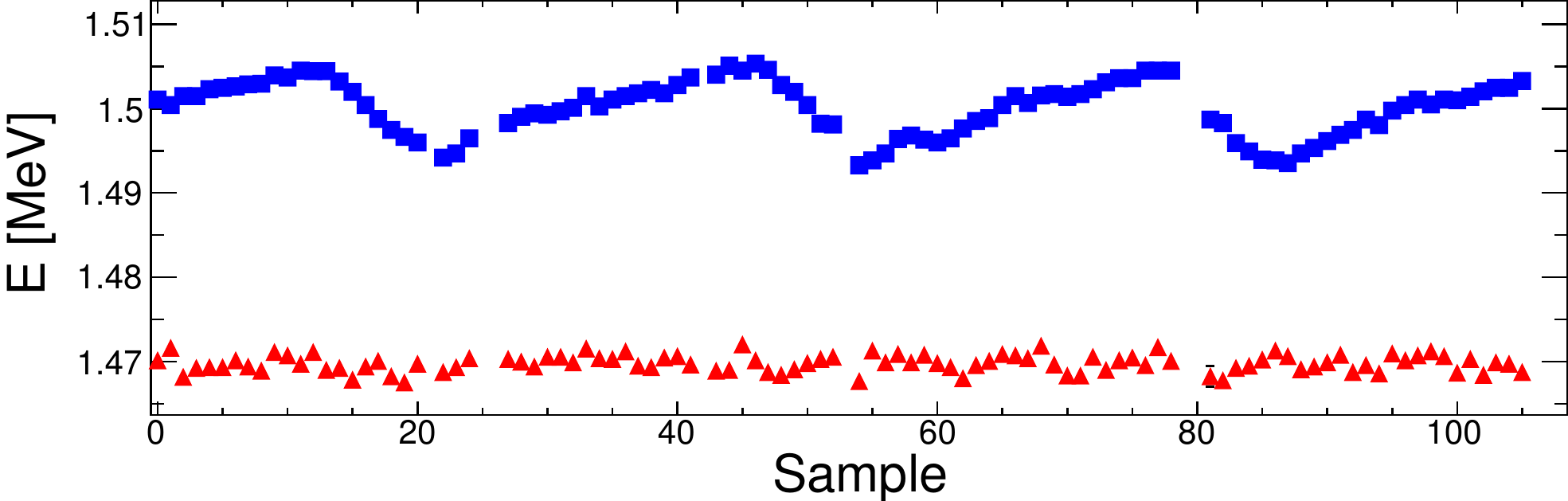}
  \caption{[color online] Correction (red triangles) of the temperature drift (blue squares) of a LaBr$_{3}$ detector over three days using data samples of around 45~min. The offset of uncorrected data reflects the strong drift since the detector calibration.}
  \label{fig:gammaWalk}
\end{figure}

LaBr$_{3}$(Ce) as a scintillator is well known to contain appreciable levels of self activity from the decay of $^{138}$La and the chemically similar $^{227}$Ac isotope~\cite{regan2015}. The former is the main source of the background with around 100~Hz per detector. The $^{138}$La nucleus decay comprises two gamma lines at 789~keV and 1436~keV from the de-excitation of the daughter nuclei $^{138}$Ce and $^{138}$Ba, respectively. The former decay is accompanied by a beta particle with an end-point energy of 258~keV while the latter gamma line is broadened due to X-rays from the electron capture escaping the crystal. These features can be well reproduced in simulation~\cite{quarati2012}. For the STELLA setup, the $^{138}$La decay pattern is implemented for all LaBr$_{3}$ assembled in the detection array and compared to packages of experimental data to obtain energy calibration correction parameters (see~\cite{heine2016} for details). The quadratic energy-response term is suppressed by $10^{-10}$ with respect to the linear term for gamma energies lower than 1.5~MeV, characterized with multiple emission lines of a $^{152}$Eu source. In the fit of experimental data to the nominal energies in the simulated self-decay spectrum, an exponential background is also taken into account. The corrections with data samples of 45~min are illustrated in Figure~\ref{fig:gammaWalk}
for the peak position of the 1.436~MeV line of the $^{138}$La decay, which is accompanied by 37~keV barium $x$-rays, for a three-day data set. The day-night cycling of  $\pm{}7$~keV (blue squares) is corrected with a precision of a few keV (red triangles). It is noteworthy to mention that the $^{138}$La decay can also be used to synchronize the time offsets of the gamma detector matrix. Here, the time stamp difference of pairs of LaBr$_{3}$ is analyzed around the 1.436~MeV line where Compton events generate unambiguous coincident signatures used to extract the time calibration parameters.

\subsection{Data Acquisition}
\label{sec:daq}

The LaBr$_{3}$ signals (FATIMA) are processed by a 1~GHz VME-based Caen$^{\textregistered}$ V1751 card that accepts external triggers and clocks to synchronize with additional devices. The QDCs are remotely controlled by MIDAS (Multi Instance Data Acquisition System), developed in the Daresbury laboratory~\cite{midas2017}.

For the charged-particle signals, commercial ABACO$^{\textregistered}$ 125~MHz $\mu$TCA compatible FMC112 cards hosting 12~ADC channels are used for digitization. In the design developed by the IPHC-SMA group at CNRS Strasbourg, two FMC112s are grouped using a FC7 AMC (Advanced Mezzanine Card)~\cite{paresi2015} with 4~GB DDR3 memory around a Xilinx Kintex7 FPGA and a communication protocol/framework based on the IPbus communication scheme. Each STELLA acquisition card provides 24 single-ended DC coupled input channels with 2~V range and a programmable DC offset correction of $\pm{}$1.25~V. The digital triggering system supports TTL compatible I/O used for the synchronization with the gamma detection system. The data readout is through a $\mu$TCA crate with Gbit ethernet communication providing remote control. The PC interface is based on the TNT corpus~\cite{arnold2006} which allows for the online analysis of single signals with a trace acquisition mode in addition to the time-stamped energy acquisition features. The Java-based software is substantially expanded for the STELLA experiment for a comprehensive handling of the DAQ with DGIC (Distributed Glibex IPbus Control), for the setup of single STELLA cards with GIC (Glibex IPbus Control), and for the merging of all data streams onto tape alongside offline analysis functionality with TAN (Tnt ANalysis) servers.

\begin{figure}[htp]
  \center
  \includegraphics[width=.9\textwidth]{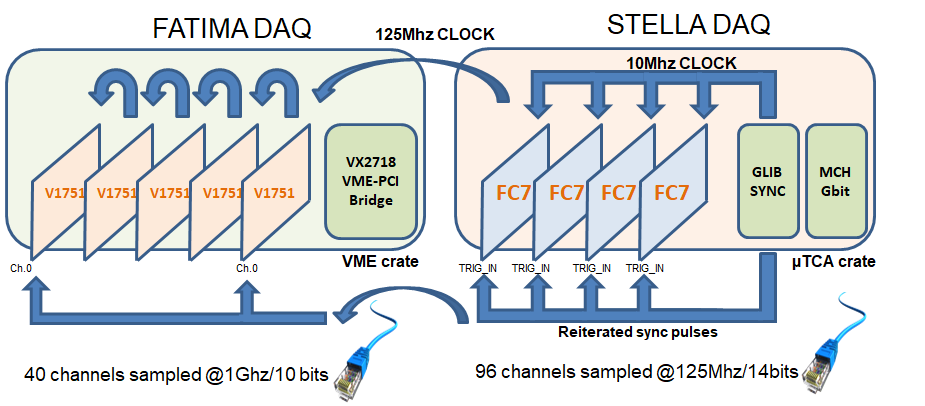}
  \caption{[color online] Time stamp synchronization of the particle (STELLA) and gamma (FATIMA) data acquisition. The clock (10~MHz) and occasional time signals are distributed by the GLIB card.}
  \label{fig:daq_sync}
\end{figure}

The time alignment of the gamma ray and particle detection is based on reference signals from a $\mu$TCA compatible GLIB (Gbit Link Interface Board) card~\cite{vichoudis2010}.  Figure~\ref{fig:daq_sync} illustrates
the distribution of a 10~MHz signal to all clocks on the respective cards for synchronization as well as the time reference signal to dedicated readout channels. The FC7 boards pass the clock to their daughter cards and distribute one 125~MHz signal from a FMC112 to the Caen V1751 modules, where it is daisy chained among the internal clocks. The time reference signals are used to determine the time-stamp offsets and to detect drifts between individual clocks.

Using four FC7 boards and five V1751 modules, 96 channels for particle and 38 channels for gamma ray detection are established. The stand-alone time-stamped trigger-less data acquisition is synchronized with reference signals from a GLIB card with a precision of a few nanoseconds.

\section{Particle-Gamma Coincidences}
\label{sec:gp_coinc}

The STELLA experiment is installed at the Androm\`{e}de accelerator~\cite{negra2016} in Orsay, \textit{France}, providing $^{12}$C beam intensities of particle-micro-ampere. Fusion reactions are measured by the coincident detection of gamma rays and light charged particles broadly following the methodology of Jiang {\em et al.}~\cite{jiang2012, jiang2018}. The excellent timing resolution of the LaBr$_{3}$(Ce) detectors used in STELLA as compared to the germanium detectors employed by Jiang {\em et al.} combined with the time-stamped data acquisition with sampling times of 8~ns for the digital triggering~\cite{arnold2006}, provides a new functionality, namely that particle-gamma timing can be used to further improve background suppression and to cleanly discriminate evaporated protons from alpha particles.

\begin{figure}[htp]
  \center
  \includegraphics[width=.7\textwidth]{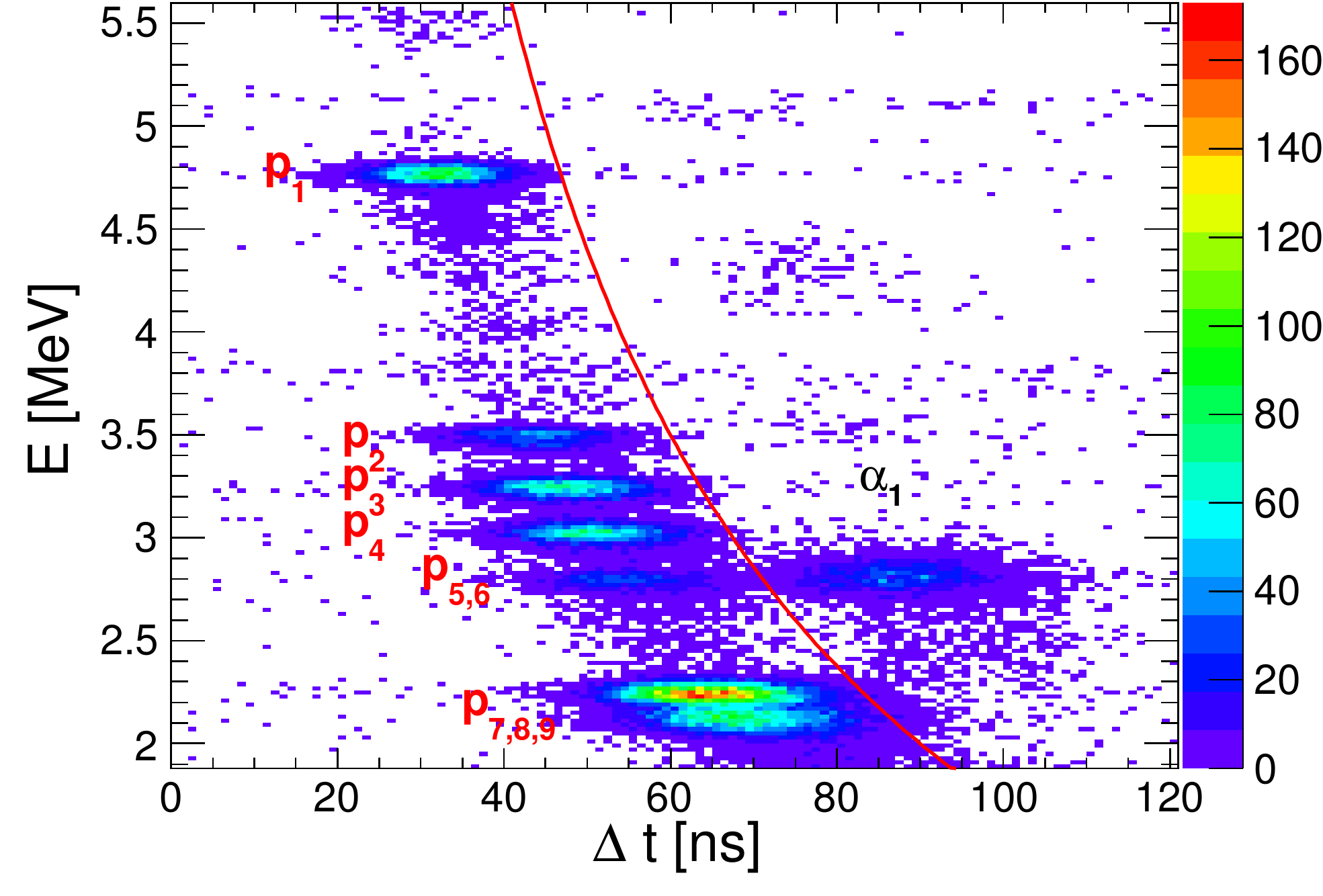}
  \caption{[color online] Time correlation of particle energies in $^{12}$C+$^{12}$C reactions with coincident gamma rays at a beam energy $E_{beam}=11$~MeV. The gamma-particle time stamp difference $\Delta{}t$ allows to distinguish protons (left of red line) and alphas (right of red line).}
  \label{fig:corr}
\end{figure}

The achievable separation is shown for the $^{12}$C+$^{12}$C reaction at a beam energy $E=11$~MeV in correlation with particle detector energy entries in Figure~\ref{fig:corr}.
The gamma-particle time stamp difference $\Delta{}t{}$ reveals contributions from protons (left of red line) and alphas (right of red line), where the time-of-flight difference cannot be resolved. The distributions are separated due to different electronic pulse shapes based on the respective energy deposition characteristics in the silicon detector substrate. These processes depend on the particle velocity that reflects in the quadratic trend of the distributions with respect to the energy of the alphas and protons~\cite{carboni2012}.

\begin{figure}[htp]
  \center
  \includegraphics[width=.7\textwidth]{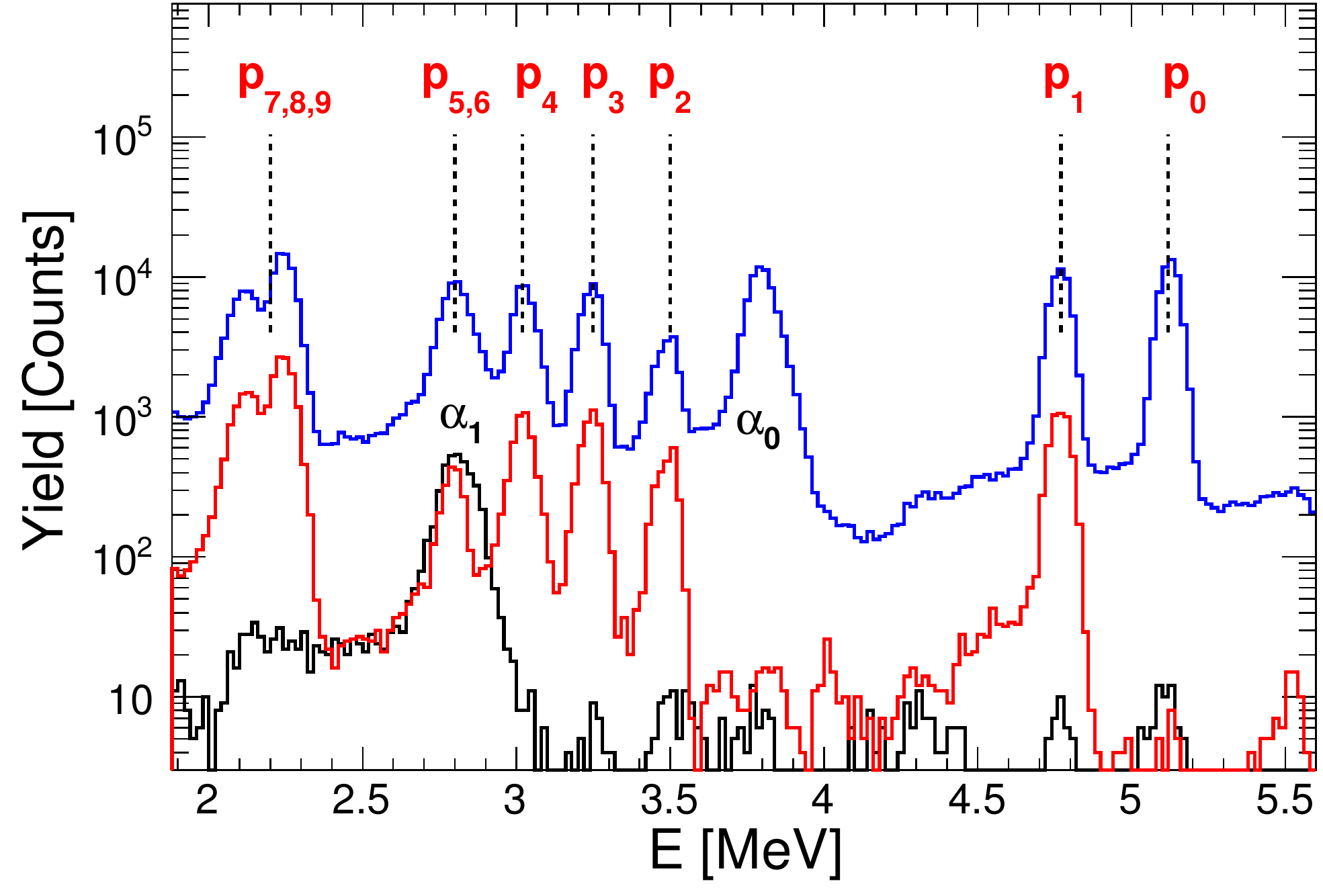}
  \caption{[color online] Particle energies at a beam energy $E_{beam}=11$~MeV where proton and alpha contributions are selected based on the timing. The distribution of all particles (blue line) can be resolved into alphas (black line) and protons (red line) coincident with gamma rays. The excitation levels $i$ of $^{20}$Ne ($\alpha_{i}$) as well as $^{23}$Na (p$_i$) are labeled and the positions of the proton energies are indicated to guide the eye.}
  \label{fig:ene_p}
\end{figure}

The selection criterion indicated by the red line in the picture is used to resolve the distributions from different particle types in the energy spectrum in Figure~\ref{fig:ene_p}.
The particle spectrum (blue line) at a beam energy $E_{beam}=11$~MeV is decomposed into the proton- (red line) and alpha-channel (black line) contributions in coincidence with gammas. Several transitions $i$ from excited states of $^{23}$Na (p$_i$) and $^{20}$Ne ($\alpha_{i}$) are labeled. Based on the timing, contributions from different particles can be resolved as can be seen for the entries around 3~MeV in the spectrum. Note that the background contamination \textit{e.g.} at the $p_{0}$ or $\alpha_{0}$ energies in the coincidence spectrum, where no associated gamma ray is emitted, is based predominantly on random coincidence with the self-activity of the LaBr$_{3}$ and can be well determined in the time domain (compare Figure~\ref{fig:corr}). Beyond this example case, the technique is also utilized to extract the relevant signals from the large overall background, essentially due to the the ubiquitous contamination of hydrogen and deuterium in the target.

\section{Summary}

The STELLA experiment has been commissioned at the Androm\`{e}de accelerator with long-run\-ning measurements of the $^{12}$C+$^{12}$C reaction using fixed targets as well as the rotating target mechanism. The rotating targets are accessible for the measurement of the thickness off the illuminated area and beside it to determine the effect of the beam exposure. The accuracy of the performance of the gamma ray detection system is guaranteed by the instant-calibration routine. It is based on the comparison of the simulated $^{138}$La decay with experimental data and has an accuracy of a few keV. Repeated alpha-source runs in the course of the campaign are utilized to ascertain the correctness of the particle detection performance.

The STELLA-FATIMA data acquisition systems are synchronized with frequently distributed time stamp pulses to dedicated readout channels. The reliability during long measurements is validated using coincident gamma-particle events from $^{12}$C fusion reactions. An enormous background reduction is achieved with the measurement of synchronous events in the gamma and particle detection system. Beyond this, reaction channels with different species of charged particles are well separated based on the timing. This guarantees a reliable measurement of deep sub-barrier partial fusion cross sections with the STELLA station.

\section{Acknowledgments}

The authors wish to thank E.~Dangelser, M. Brucker, H.~Friedmann, J.-N.~Grapton, H.~Kocher, C.~Mathieu, C.~Rues\-cas, C.~Schwab, D.~Thomas, S.~Veeramootoo as well as F.~Agnese, O.~Clausse, L.~Gross, M.~Imhoff, and C.~Wabnitz (IPHC-CNRS, Strasbourg, \textit{France}) for their valuable contribution with the construction of the measurement station. We are fortunate to benefit from valuable discussions with J.~Faerber (IPCMS, Strasbourg, \textit{France}) to establish the ultra-high vacuum . Furthermore, we thank M.~Lorrigiola (LNL, Padova, \textit{Italy}) and G.~Fr\'{e}mont (GANIL, Caen, \textit{France}) for the excellent preparation of the reaction targets. The STELLA construction is funded by the University of Strasbourg IdEX program and CNRS Strasbourg. The Androm\`{e}de facility (ANR-10-EQPX-23) is funded by the program for future investment EQUIPEX. This work is also partially supported by the UK Science and Technology Facility Council (UK) \textit{via} grants ST/L005743/1 and ST/P005314/1. P.H.~Regan also acknowledges support from the UK National Measurement Office.

\section*{References}

\bibliographystyle{elsarticle-num}
\bibliography{myBibFile}

\end{document}

%% file: stella_collab.tex
\author[iphc,cnrs]{M.~Heine~\corref{corrAuthor}}

\author[iphc,cnrs,usias]{S.~Courtin}
\author[iphc,cnrs]{G.~Fruet}
\author[york]{D.G.~Jenkins}
\author[york]{L.~Morris}
\author[iphc,cnrs,usias]{D.~Montanari}
\author[surrey]{M.~Rudigier}
\author[ipn]{P.~Adsley}
\author[iphc,cnrs]{D.~Curien}
\author[ipn]{S.~Della Negra}
\author[ipn]{J.~Lesrel}

\author[iphc,cnrs]{C.~Beck}
\author[iphc,cnrs]{L.~Charles}
\author[iphc,cnrs]{P.~Den\'{e}}
\author[iphc,cnrs]{F.~Haas}
\author[ipn]{F.~Hammache}
\author[iphc,cnrs]{G.~Heitz}
\author[iphc,cnrs]{M.~Krauth}
\author[ipn]{A.~Meyer}
\author[surrey]{Zs. Podoly\'{a}k}
\author[surrey,tedd]{P.H.~Regan}
\author[iphc,cnrs]{M.~Richer}
\author[ipn]{N.~de S\'{e}r\'{e}ville}
\author[ganil]{C.~Stodel}

\address[iphc]{IPHC, Universit\'{e} de Strasbourg, Strasbourg, F-67037 (France)}
\address[cnrs]{CNRS, UMR7178, Strasbourg, F-67037 (France)}
\address[usias]{USIAS/Universit\'{e} de Strasbourg, Strasbourg, F-67083 (France)}
\address[york]{University of York, York, YO10 5DD (UK)}
\address[ipn]{IPN d'Orsay, UMR8608, CNRS/IN2P3, PSUD 11, Orsay, F-91406, (France)}
\address[surrey]{Department of Physics, University of Surrey, Guildford, GU2 7XH (UK)}
\address[tedd]{National Physical Laboratory, Teddington, Middlesex, TW11 0LW (UK)}
\address[ganil]{GANIL, CEA/DSM-CNRS/IN2P3, Caen, F-14076 (France)}

\cortext[corrAuthor]{\href{mailto:marcel.heine@iphc.cnrs.fr}{marcel.heine@iphc.cnrs.fr}}